\documentclass[aps,prl,twocolumn,groupeaddress,showpacs,psfig]{revtex4}
\usepackage{psfig}

\begin{document}

\title{Vibrational Enhancement of the Effective Donor - Acceptor Coupling}

\author{\sc M. Lazrek$^{1,2}$, D. J. Bicout$^{1,3}$, S. Jaziri$^{2}$ and E. Kats$^{1,4}$}

\affiliation{$^1$Institut Laue-Langevin, 6 rue Jules Horowitz,
BP 156, Grenoble, France \\
$^{2}$ Laboratoire de Physique de la Mati\`{e}re Condens\'ee, 
Facult\'e des Sciences de Bizerte, 7021 Jarzouna Bizerte, Tunisie\\
$^{3}$ Biomathematics and Epidemiology, ENVL - TIMC, 
B.P. 83, 69280 Marcy l'Etoile, France\\
$^{4}$ L. D. Landau Institute for Theoretical Physics,
RAS, 117940 GSP-1, Moscow, Russia}

\date{\today}

\begin{abstract}
The paper deals with a simple three sites model for charge transfer phenomena in 
an one-dimensional donor (D) - bridge (B) - acceptor (A) system coupled with 
vibrational dynamics of the B site. It is found that in a certain range of parameters 
the vibrational coupling leads to an enhancement of the effective donor - acceptor 
electronic coupling as a result of the formation of the polaron on the B site. This  
enhancement of the charge transfer efficiency is maximum at the resonance, 
where the effective energy of the fluctuating B site coincides with the 
donor (acceptor) energy.

\end{abstract}


\maketitle

Molecular electronics is progressing so rapidly that it is now possible to 
do measurements and assembly at the level of individual or few molecules \cite{BS00,PB00}. 
Charge transport is known to occur in a wide range of linear chain molecules 
including the DNA double strand molecules. For DNA, it is believed that the charge 
transport phenomenom is involved in the protection of the DNA encoded 
information against the oxidative damage \cite{HE00}. As the DNA molecule is 
essentially a dynamic structure on the time scale of charge transport, one 
expects that vibrational dynamics to play an important role for DNA electronics, 
and, in general, for any property of biological molecules because biological 
functions of life are associated with molecular motions but not to the static or 
dead structure (i.e., equilibrium positions of all atoms). 

In this paper we are interested in an one-dimensional DNA wire or bridge (B) 
connecting a donor (D) and an acceptor (A) sites. Usually, the bridge consists of 
$N$ sites with one state per site (see the abundant literature devoted to 
this issue in the Refs.\cite{AS96,EK93,DH97,HB97,AS97,FT98,FS99,WF00,GA01,
SS01,BB02,NR03,TH02,JB99,BVK05,r6,r7,r8,r9,KL02}), and theoretical analysis of this 
problem requires to solve a system of $N+2$ non-linear coupled equations. 
Unfortunately, such a problem cannot be solved analytically for $N \gg 1$, and 
we have to recourse to numerical solution. However, many insights and essential 
features of the dynamics can already be gained and captured by studying a 
simple three sites: D - B - A. Generally speaking, the interaction between the 
donor and acceptor involves all states of the bridging subsystem. This bridge 
mediated interaction can be characterized, under certain conditions, by a single 
energy dependent parameter - effective coupling - which plays the key role in 
the charge transfer. For small systems the phase coherence of charges is maintained 
over the entire system, and the quantum effects are crucial in determining the 
system properties. On the contrast, for the long $N \gg 1$ bridges, fast 
relaxation processes result in a strong dephasing between charges in the system. 
Therefore, this leads to a rapid falloff of the off-diagonal elements of the 
density matrix such that the diagonal elements can described by a set of 
kinetic equations \cite{JB99,BK02}.

Our concern in this paper is to investigate a Hamiltonian model describing the 
D - A coupling under in the presence of dynamic structural fluctuations. Such local 
fluctuations, including local vibrations, twist motions, radial deformations  
and hydrogen-bond stretching or opening, are known to strongly influence charge 
transfer in DNA molecules \cite{BG00,OR02,H00,L00}. For simplicity, we consider 
a three sites D - B - A system where electronic degree of freedom is coupled to 
an effective local vibrational degree of freedom.

Let us assume that initially the charge is entirely localized on the
donor site with energy $\epsilon $. Then owing to the non-zero overlapping
integrals of the electronic wave functions between the two neighboring sites, the 
tunneling of the charge takes place from the donor to the acceptor site with 
the same energy. Denoting by $\{|d\!>,|b\!>,|a\!>\}$ the localized states
on the the donor, bridge and acceptor, respectively, the Hamiltonian of the 
bridge - mediated charge transfer between the donor and acceptor acquires the form, 
\begin{eqnarray}
\label{l1}
H  =  H_{\rm e} +\frac{p^2}{2m}+\frac{m\omega ^2r^2}{2}\,+\,
kr\,|b\!><\!b| \, ,
\end{eqnarray}
in which the bare electronic part reads as,
\begin{eqnarray}
\label{l111}
H_{\rm e}  =  \varepsilon\,\left[|d\!><\!d|+|a\!><\!a|\right]
+\varepsilon_b\,|b\!><\!b|  \\
 +  v_{db}\left[|d\!><\!b|+|b\!><\!d|\right]\nonumber \\
 +  v_{ba}\left[|b\!><\!a|+|a\!><\!b|\right]\nonumber \, ,
\end{eqnarray}
where $\epsilon$ is the one-site energy of the donor and acceptor, $\epsilon_b$ 
the one-site energy of the bridge, $m$ ($m \simeq 300 amu$) is the mass of the 
bridge base pair, $r$ is its radial displacement in the localized vibrational 
mode with frequency $\omega $, the momentum $p$ is conjugated to $r$, and $k$ 
is the electron-localized vibration mode coupling constant. The localized bridge 
mode can be treated classically since the corresponding vibrational displacement 
amplitude is larger than zero point quantum fluctuations for characteristic DNA 
parameters \cite{BG00,OR02,H00,L00} (see also the following). 

The frequency of typical vibrations in DNA ranges  
$\omega \simeq 10^{11} - 10^{12} s^{-1}$, and the scale of electronic overlap 
integrals between base pairs in DNA is $v=\sqrt{2(v_{db}^2+v_{ba}^2)}\simeq 0.2\,$eV 
leading to an electronic characteristic frequency, 
$v/\hbar\simeq 3\times 10^{14} s^{-1}$. As a consequence of 
the small (adiabatic) parameter, $v\omega/\hbar\ll 1$, 
the slow vibrational and fast electronic motions can be decoupled. Therefore,  
to solve the problem of bridge-mediated charge transfer between the donor and the 
acceptor, we employ the adiabatic procedure to eliminate the slow 
vibrational motions and derive an effective Hamiltonian for fast electronic motions.
To proceed, we take the wave function of the charge in the form, 
$
|\Psi(t)\!>=c_d(t)|d\!>+c_b(t)|b\!>+c_a(t)|a\!> \, ,
$
where $c_n(t)$ are the time dependent amplitudes of the probability to have
the charge at the $n$th site. From the Hamiltonian (\ref{l1}) we arrive
at the 
following equations of motion for the quantum amplitude $c_n(t)$,
\begin{eqnarray}
\label{l2}
i\hbar\,\frac{d}{dt}\left(
\begin{array}{c}
c_d\\c_b\\c_a
\end{array}\right)=\left(\begin{array}{ccc}
\varepsilon & v_{db} & 0 \\
v_{db} & \varepsilon_b+kr & v_{ba}\\
0 & v_{ba} & \varepsilon
\end{array}\right)\,\left(
\begin{array}{c}
c_d\\c_b\\c_a
\end{array}\right)\:,
\label{dtel}
\end{eqnarray} 
and for the classical dynamic mode,
\begin{eqnarray}
\label{l3}
m\frac{d^2r}{dt^2}=-m\omega^2r-k\,|c_b|^2\:.
\end{eqnarray}
Next, we seek for stationary solutions of the form $c_n(t)=c_n\,{\rm e}^{-iEt/\hbar}$
oscillating with frequency $E/\hbar$. To work with dimensionless quantities, 
we use from now the dimensionless variables $u$, $\sigma$ and $\kappa$ defined 
in Table~\ref{tb1}. Using $c_n(t)$ in the equations of motions, we find that the 
stationary bridge displacement is $r=-\kappa\,c_b^2$ and the stationary 
probability distribution is, 
\begin{eqnarray}
\label{prob}
\left\{\begin{array}{l}
c_d^2=\eta^2c_a^2 \\
c_b^2=2u^2/(1+2u^2) \\
c_a^2=1/\left[(1+\eta^2)(1+2u^2)\right]
\end{array}\right.
\end{eqnarray}
where the eigen-energy $u$ satisfies the characteristic equation, 
\begin{equation}
4u^4+4(\kappa^2-\sigma)u^3-2\sigma u-1=0\:.
\label{root}
\end{equation}
Solutions of this equation provides the ground $u_g$ plus one or three (depending 
upon values of $\kappa$ and $\sigma$) excited energies of the ``polaron'', 
i.e. the state created by the charge coupling with the DNA structural deformation.
\begin{table}
\caption{Definition of dimensionless variables.}
\begin{tabular}{ll}  \hline\hline
definition        & variable\\ \hline
energy scale      & $v=\sqrt{2(v_{db}^2+v_{ba}^2)}$ \\
length scale      & $\xi=\sqrt{v/m\omega^2}$ \\
coupling asymmetry  & $\eta=v_{db}/v_{ba}$  \\
dimensionless polaron energy   & $u=(E-\varepsilon)/v$ \\
dimensionless energy barrier & $\sigma=(\varepsilon_b-\varepsilon)/v$ \\
reduced electron-vibration coupling & $\kappa=k\xi/v$ \\ \hline \hline
\end{tabular}
\vspace{0.5cm}
\label{tb1}
\end{table}
These stationary polaron solutions allows us to eliminate from (\ref{l1}) the 
structural deformation $r$ and to obtain the effective Hamiltonian for the 
charge transfer as,
\begin{eqnarray}
\label{l5}
H_{\rm eff} & = & \frac{1}{v}\left(H_{\rm e}-\varepsilon{\hat 1}\right)=
\Delta(u)\,|b\!><\!b| \nonumber \\
& + & \frac{\eta}{\sqrt{2(1+\eta^2)}}\,\left[|d\!><\!b|+|b\!><\!d|\right] \\
& + & \frac{1}{\sqrt{2(1+\eta^2)}}\,\left[|b\!><\!a|+|a\!><\!b|\right]
\nonumber \:,
\end{eqnarray}
where, $\Delta(u)=\sigma-\kappa^2c_b^2(u)$, is the renormalized effective 
energy of the bridge due to electron - vibration coupling, and $\hat 1$ is 
the unit matrix. To calculate the effective charge coupling between the donor 
and the acceptor, we have to solve first the time dependent problem to determine 
the probability of charge transfer defined as, 
$P_{d\rightarrow a}(t)=|<\!a|\Psi(t)\!>|^2$, where
$|\Psi(t)\!>$ is the solution of the Schr\"odinger equation,
$$
i\hbar\,\frac{d|\Psi(t)}{d t}\!> = H_{\rm eff}|\Psi(t)\!>\, 
$$
with the initial condition,
$|\Psi(0)\!>=|d\!>$. It is easy to show that, when $t\rightarrow\infty$ we have 
$P_{d\rightarrow a}(t)\simeq k_{da}t$, where the charge transfer rate
$k_{da}$ is given by the Fermi Golden rule,
\begin{eqnarray}
\label{l6}
k_{da}=\frac{2\pi}{\hbar}|H_{da}|^2\times\,\left\{\begin{array}{ccl}
\delta(u_2-u_1) & ; & \sigma\geq \kappa^2c_b^2\\
\\
\delta(u_3-u_1) & ; & \sigma\leq \kappa^2c_b^2
\end{array}\right.
\,
\end{eqnarray}
with the (dimensionless) effective donor - acceptor coupling \cite{BVK05}, 
\begin{eqnarray}
& & |H_{da}|^2=\frac{\eta^2}{4(1+\eta^2)^2}\nonumber \\
& & \times\,\left\{\begin{array}{ccl}
1/\left[(u_3-u_1)(u_3-u_2)\right] & ; & \sigma\geq \kappa^2c_b^2\\
\\
1/\left[(u_1-u_2)(u_3-u_2)\right] & ; & \sigma\leq \kappa^2c_b^2
\end{array}\right.
\label{ecp}
\end{eqnarray}
where $u_i$ are the eigen-energies of $H_{\rm eff}$ given by $u_1=0$ and 
$2u_{2,3}=\Delta\mp\sqrt{\Delta^2+2}$. Finally, we end up with the 
effective D-A coupling given by, 
\begin{eqnarray}
\label{hda}
& & |H_{da}(u,\eta,\sigma,\kappa)|^2=\frac{\eta^2}{2(1+\eta^2)^2\,
\sqrt{\Delta^2+2}}\nonumber \\
& & \times\,
\frac{1}{\left[|\Delta|+\sqrt{\Delta^2+2}\right]}\:,
\end{eqnarray}
and the ratio $\rho$, allowing to measure the effect of vibrations on the 
D-A coupling, reads:
\begin{eqnarray}
& & \rho(u,\sigma,\kappa)=\frac{|H_{da}(u,\eta,\sigma,\kappa)|^2}
{|H_{da}(u,\eta,\sigma,0)|^2}\nonumber \\
& & =\left[
\frac{|\sigma|+\sqrt{\sigma^2+2}}{|\Delta|+\sqrt{\Delta^2+2}}\right]
\,\times\,\left[\frac{\sigma^2+2}{\Delta^2+2}\right]^{1/2}\:.
\label{rho}
\end{eqnarray}
The expressions (\ref{hda}) and (\ref{rho}) are the main results of 
this paper. They provide close formulas for evaluating how the 
dynamical disorder affects the effective donor-acceptor coupling in
various situations. As a direct applications of our main findings, 
we consider the following illustrative examples. 

{\bf Charge density versus polaron energy:}

\begin{figure}[ht]
\vspace{0.3cm}
\centerline{\psfig{figure=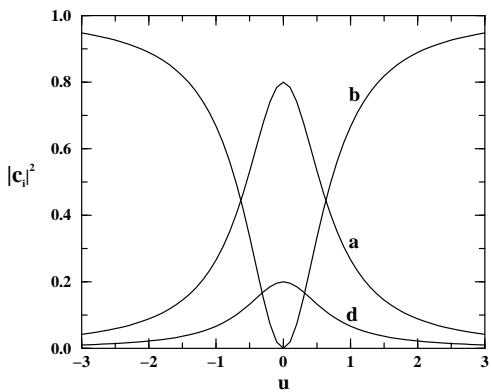,width=2.5in,angle=0}}
\vspace{-0.25cm} \caption{Stationary charge density in Eq.(\ref{prob}) 
versus energy for the coupling asymmetry $\eta=0.5$. The quoted letters 
``d'', ``b'', and ``a''stand for the donor, bridge and acceptor, 
respectively.} \label{fig1}
\end{figure}

It results from Eq.(\ref{prob}) that the charge densities on the donor, 
bridge and acceptor are even functions of the polaron energy $u$. For all $u$, 
the ratio of the density of the donor to that of the acceptor is equal the 
square of the asymmetry energy $\eta$ (see the Table ~\ref{tb1}).
For $u=0$, there is no charge on the bridge 
site, and the charge density is distributed between the donor and 
acceptor sites in proportion of $\eta$. In contrast, for the limits
of very high (or low) polaron energy when $|u|$ gets larger, 
the charge density decreases considerably
on the donor and acceptor sites while it gets higher on the bridge site 
leading hence to small charge transfer efficiency. These features 
are illustrated in Fig.~\ref{fig1} for the energy asymmetry parameter  $\eta = 0.5$. 
Likewise, at the resonance $u_m$ where the renormalized effective 
energy of the bridge is equal to zero (the donor/acceptor energy),
\begin{equation}
\label{resn}
\Delta(u_m)=0\:\:\Longrightarrow\:\:
u_m=\pm\frac{\sqrt{\sigma}}{\sqrt{2(\kappa^2-\sigma)}}\:,
\end{equation}
the distribution of charge density reduces to, 
\begin{eqnarray}
\label{probm}
\left\{\begin{array}{l}
c_d^2(u_m)=\eta^2c_a^2(u_m) \\
c_b^2(u_m)=\sigma/\kappa^2 \\
c_a^2(u_m)=[1-c_b^2(u_m)]/(1+\eta^2)
\end{array}\right.
\end{eqnarray}
At this resonance point, the charge density on the bridge site decreases 
either upon approaching to the bare resonance for $\sigma\rightarrow 0$ or 
by increasing the electron-vibration coupling parameter $\kappa$ above $\sqrt{\sigma}$. 

{\bf Effective D-A coupling versus $u$:}

The effective D-A coupling scales as, $|H_{da}|\sim 1/\Delta$ for $\Delta\gg 1$. 
However, at the resonance $\Delta=0$ defined in Eq.(\ref{resn}), the  
$|H_{da}(u,\eta,\sigma,\kappa)|^2$, and thus the ratio $\rho(u,\sigma,\kappa)$, 
attain their maxima given by, 
\begin{eqnarray}
|H_{da}(u_m,\sigma,\kappa)|^2=\frac{\eta^2}{4(1+\eta^2)^2}\:,
\end{eqnarray}
and, 
\begin{eqnarray}
\rho(u_m,\sigma,\kappa)=\frac{\left(\sigma+\sqrt{\sigma^2+2}\right)
\sqrt{\sigma^2+2}}{2}\:.
\label{rhm}
\end{eqnarray}  
Simple inspection of this equation shows
two characteristic features. First, at the resonance the effective 
D-A coupling is enhanced by the coupling to
structural dynamics, and second, as it is illustrated 
in the Fig.~\ref{fig2}, the enhancement factor $\rho$ of the 
effective D-A coupling due to electron-vibration interactions increases 
with energy barrier $\sigma$. 
To rationalize these observations in terms of the polaron energy and electron-vibration 
coupling, we have depicted in Fig.~\ref{fig2} the ratio $\rho(u,\sigma,\kappa)$ 
as a function of $u$ for increasing values of $\sigma$ and $\kappa$. Two different 
regimes can be distinguished: 
\begin{itemize}
\item $\kappa< \sqrt{\sigma}$: below the resonance value, the effective D-A 
coupling is a monotonic increasing function of the polaron energy $|u|$. 

\item $\kappa \geq \sqrt{\sigma}$: the effective D-A coupling increases for 
$|u|<|u_m|$, attains its maximum at the resonance $|u|=|u_m|$, and decreases 
for $|u|>|u_m|$. As a consequence of $|H_{da}(u,\sigma,0)|\sim 1/\sigma$, 
both the maximum of $\rho $ at the resonance and its limit at high $|u|$
increase with $\sigma$. 
\end{itemize}
\begin{figure}[ht]
\vspace{0.3cm} \centerline{
\psfig{figure=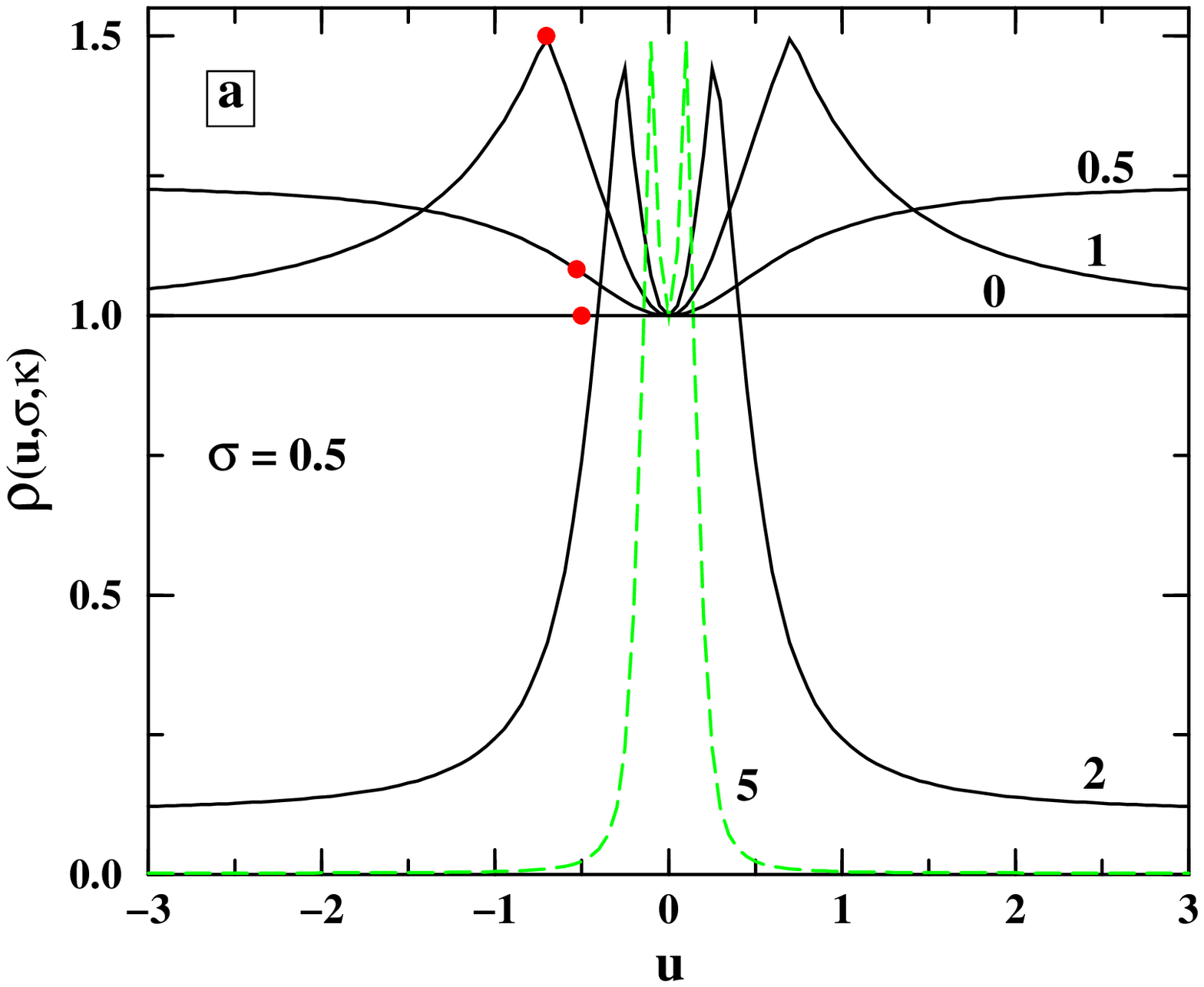,width=2.5in,angle=0}}
\end{figure}
\begin{figure}[ht]
\vspace{0.3cm}\centerline{
\psfig{figure=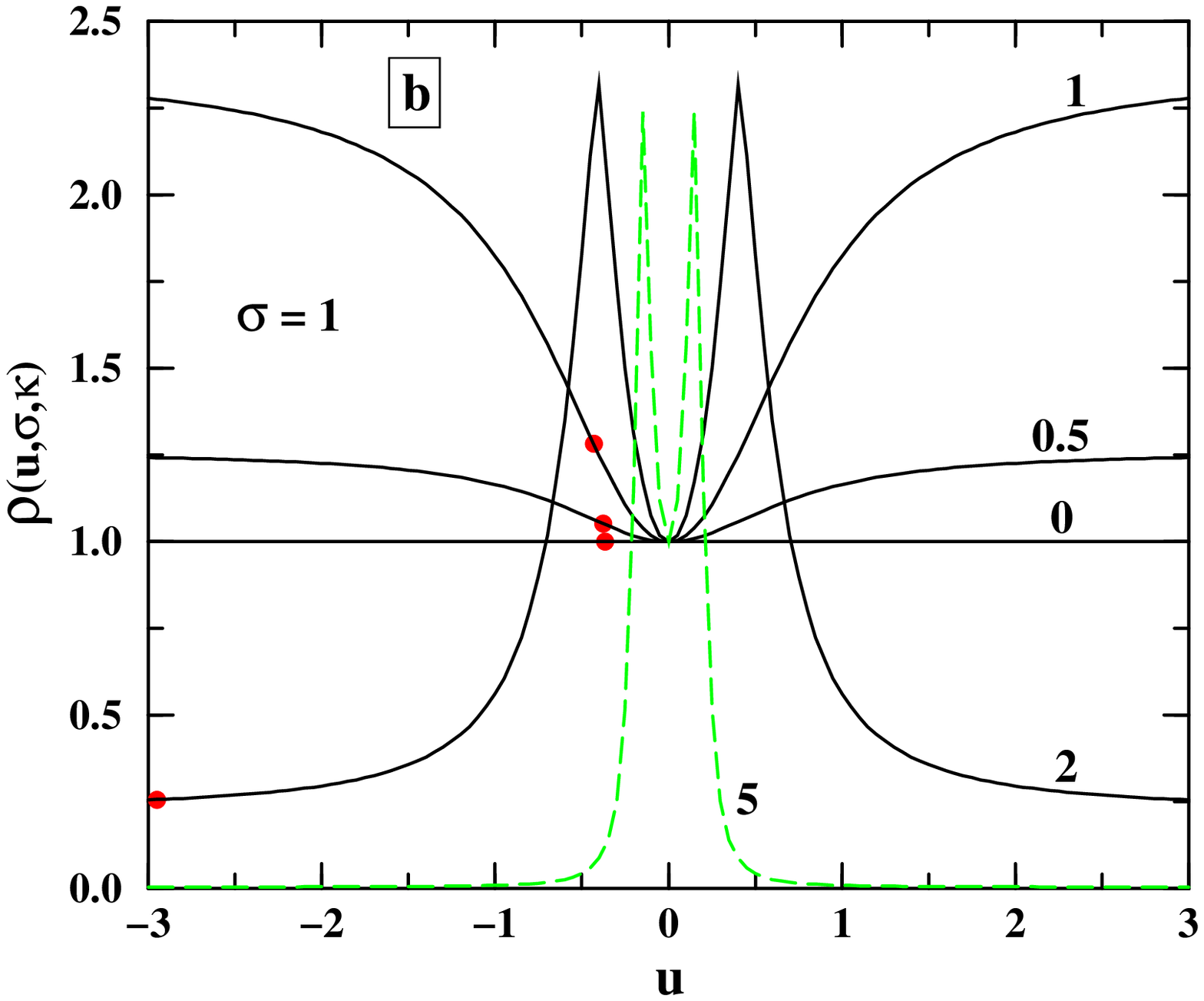,width=2.5in,angle=0}}
\end{figure}
\begin{figure}
\vspace{0.3cm}\centerline{
\psfig{figure=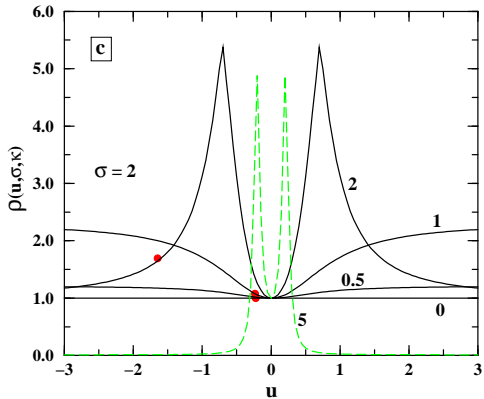,width=2.5in,angle=0}} \vspace{-0.25cm}
\caption{Effective coupling ratio in Eq.(\ref{rho}) as a
function of the energy. Filled circles correspond to
$\rho(u_g,\sigma,\kappa)$ at the ground state energy $u_g$. The
quoted numbers correspond to the electron-vibration coupling values,
i.e., $\kappa=0,\,0.5,\,1,\,2,\,5$.} \label{fig2}
\end{figure}

{\bf Effective D-A coupling versus $\kappa $:}

\begin{figure}[ht]
\vspace{0.3cm}
\centerline{\psfig{figure=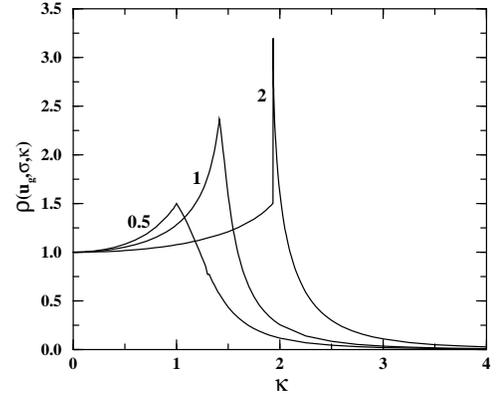,width=2.5in,angle=0}}
\vspace{-0.25cm} \caption{Effective coupling ratio in
Eq.(\ref{rho}) at the ground state energy $u_g$ as a function of
the electron-vibration coupling. The quoted numbers correspond to the
energy barrier values, i.e., $\sigma=0.5,\,1,\,2$.} \label{fig3}
\end{figure}

As we have discussed above and illustrated in Fig.~\ref{fig2}, the 
electron-vibration may lead to an increase or decrease of the effective 
D-A coupling depending on the value of $|u|$ and the regime of $\sigma$. 
Similarly, Fig.~\ref{fig3} displays the enhancement factor $\rho $ at 
the polaron ground state as a function of the electron-vibration coupling 
$\kappa $. It is clear that there is a certain threshold value 
$\kappa_c(u_g)$ below which the electron-vibration coupling leads to  
enhancement of the effective D-A coupling, and above which the effective 
D-A coupling is drastically reduced affecting hence the charge transfer 
efficiency. 

As above, two different regimes can be distinguished: 
\begin{itemize}
\item $\sigma\leq 1$: Equation (\ref{root}) has two distinct roots corresponding 
to the ground and excited states, respectively. The maximum enhancement, 
$\rho(u_m,\sigma,\kappa_c)$ given by Eq.(\ref{rhm}), is attained at the resonance 
$\Delta=0$, where $u_g=-\sqrt{2}/2$ (and the excited state, $-u_g$) and 
$\kappa_c(u_g)=\sqrt{2\sigma}$, obtained from the combination of Eqs.(\ref{root}) 
and (\ref{resn}). In this regime, the ground state coincide with the resonant energy, 
$u_g=u_m$. 

\item $\sigma>1$: There is an interval, $\kappa_c(u_g)\leq \kappa\leq\kappa_{max}$, 
within which Eq.(\ref{root}) admits four distinct roots (the lowest one 
corresponding to the ground state) and out of which it has two distinct roots. 
In this case, the ground state is no longer resonant, $u_g\neq u_m$, but two 
excited states coincide with the resonant energies given by $u_m=\pm\sqrt{2}/2$. 
As a result, the maximum enhancement $\rho(u_g,\sigma,\kappa)<\rho(u_m,\sigma,\kappa_c)$ 
as $\Delta(u_g)\neq 0$. For instance, for $\sigma=2$, the four distinct roots 
interval is $1.9336\leq \kappa\leq 2.175$ with $\kappa_c=1.9336$ and the maximum 
enhancement $\rho(u_g,\sigma,\kappa_c)=3.198$ is obtained for $u_g=-1.097$. At the 
resonance $\Delta=0$ for $\sigma=2$, we have $\kappa=2$, $u_g=-1.707$, 
$\Delta(u_g)=-\sqrt{2}$ and $\rho(u_g,\sigma,\kappa)=1.596$. 
\end{itemize}

In summary, we have shown that the electronic coupling with the vibration dynamics 
of the bridge results in a formation of a polaron that may, under certain conditions, 
leads to an enhancement of the charge transfer efficency. Figures~\ref{fig2} and 
\ref{fig3} show that the enhancement factor $\rho $ is greater than one for a wide 
range of the energy barrier $sigma$ and the electron-vibration coupling $\kappa$. 
These findings are very suggestive for the issue of charge transport assisted 
by structural dynamical along the DNA chain. To study the basic mechanism of 
vibration enhancement of charge transport we have focused in this work
on the simple three sites model with a single harmonic structural dynamic mode 
(reaction coordinate). Meanwhile, the method employed in this work is not 
limited to this model and the extension of the theory to several sites and 
anharmonic reaction coordinates (see e.g., Refs.\cite{H00,L00}), and several 
resonance states can be handled within the framework developed in Ref.~\cite{BVK05}. 
Nevertheless, further theoretical studies need to be conduct along the ideas 
outlined above in order to gain a better understanding of charge transport 
properties in biological systems and technological applications of significant 
importance.


\end{document}